\documentclass[aps,twocolumn,floatfix,nofootinbib]{revtex4-2}
\usepackage{graphicx}
\usepackage{dcolumn}
\usepackage{bm}
\usepackage{times}
\usepackage{graphics}
\usepackage{bm}
\usepackage{subfigure}
\usepackage{graphicx}
\usepackage{amsthm}
\usepackage{notes2bib}
\usepackage{amssymb}
\usepackage{url}
\usepackage{enumerate}
\usepackage{epstopdf}
\usepackage{mathtools}          
\usepackage{setspace}
\usepackage[usenames,dvipsnames,svgnames,table]{xcolor}
\usepackage{color}

\newcommand{\allblack}{\color{black}{}}

\begin{document}
\preprint{APS/123-QED}
\title{Data-driven ODE modeling of the high-frequency complex dynamics via a low-frequency dynamics
model}
\author{Natsuki Tsutsumi}
\affiliation{Faculty of Commerce and Management, Hitotsubashi University, Tokyo 186-8601, Japan}
\author{Kengo Nakai}
\affiliation{The Graduate School of Environment, Life, Natural Science and Technology, Okayama University, Okayama 700-0082, Japan}
\author{Yoshitaka Saiki}
\affiliation{Graduate School of Business Administration, Hitotsubashi University, Tokyo 186-8601, Japan}
\date{\today}

\begin{abstract}
In our previous paper [N. Tsutsumi, K. Nakai and Y. Saiki, Chaos 32, 091101 (2022)],
we proposed a method for constructing a system of differential equations of chaotic behavior from only observable deterministic time series, which we call the radial function-based regression (RfR) method. 
However, when the targeted variable's behavior is rather complex, the direct application of the RfR method does not function well. In this study, we propose a novel method of modeling such dynamics, including the high-frequency intermittent behavior of a fluid flow, by considering another variable
(base variable) 
showing relatively simple, less intermittent behavior. We construct an autonomous joint model composed of two parts: the first is an autonomous system 
of a 
base variable,
and the other concerns the targeted variable being affected by a term involving the 
base variable 
to demonstrate complex dynamics. 
The constructed joint model succeeded in 
not only inferring a short trajectory but also reconstructing chaotic sets and statistical properties obtained from a long trajectory such as the density distributions of the actual dynamics.
\end{abstract}

\maketitle

{\it Introduction.}
We are often eager to identify a dynamical system model that generates observable time series data. 
Many systems that describe physical phenomena have been found since the finding of the equation of motion.
In decades, various methods have been proposed that estimate a dynamical system describing observed data combining physical knowledge and machine learning techniques.
Chorin et al~\cite{chorin2015} proposed a method to 
describe unknown partial dynamics that are not described by equations of physical systems.
Physics-informed machine learning methods~\cite{karniadakis2021,raissi2019,greydanus2019} 
construct a model 
by supposing physical conditions that a system generating time series is expected to have.
Sometimes, we do not have any physical knowledge of a system that generates time series.
Several approaches have been proposed concerning modeling dynamics using machine learning from given time series data without physical knowledge.
The dynamic mode decomposition~\cite{schmid2010} is a method to derive macroscopic features and estimate a dynamical system of the variables.
Reservoir computing~\cite{Pathak_2017,nakai_2018,nakai_2020,kobayashi2021} is a recurrent neural network 
with low computational cost.
These methods usually model a continuous dynamical system as discrete-time, not differential equations.
Note that a model constructed using reservoir computing is commonly a discrete-time system, 
but there are some types of reservoir computing using differential equations to describe the dynamics of the reservoir states~\cite{Lu_2018}.

In some studies, a system of ordinary differential equations (ODEs) is constructed to describe the dynamics of deterministic time series.
Neural ODE~\cite{chen2018} recently attracted much attention for using a neural network to model ODE. 
The method based on the theory of the Koopman operator~\cite{lin2021, mezic2013, berry2015} projects the phase space to an infinite dimensional space to construct a model as a linear system.
There are studies to make a simple and understandable model of ODEs in a finite dimensional space without a neural network.
Several studies~\cite{baake1992,wang2011,brunton16, champion19} have estimated a system of ODEs from the time series of all variables of the background system.
Other investigations~\cite{gouesbet91,gouesbet92,gouesbet97} have derived a system of ODEs using an observable variable and its time derivatives as the model variables.
In these studies, prior knowledge of the background dynamics was used to choose basis functions for the regression; however, this method is not appropriate for some practical purposes~\footnote{In some situations, prior knowledge of the background dynamics can be obtained and the modeling with the knowledge is useful~\cite{pathak2018, wikner2020}.}.

Recently, we proposed a simple method for constructing a system of ODEs for chaotic behavior based on regression using only observable 
deterministic
scalar time series data~\cite{tsutsumi22,tsutsumi23},
referred to as the radial function-based regression (RfR) method.
The RfR method constructs a model describing the dynamics of a time series in a space made only from understandable variables without background knowledge.
For the regression, we employ the delay coordinate of the observable variable and spatially localized radial basis functions in addition to polynomial basis functions. 
The introduction of the localized functions leads to higher computational costs than that of some other methods such as reservoir computing. 
The method's forecasting capability is not limited to a short time series but also to a density distribution created using a long time series.
We can generate a long trajectory 
of the model using the Stagger-and-Step method~\cite{sweet_2001c} when the appropriate model trajectory is realized on a chaotic set 
of the model that is not an attractor.
This RfR method functions
even when observable variables are obtained  
from an infinite-dimensional system such as partial differential equations and delay differential equations~\cite{tsutsumi23}.

Regardless of the modeling method, some types of complex time series governed by dynamics with multiscale structures and/or high Lyapunov dimensions are difficult to model.
In the case of a fluid flow, 
direct modeling of high-frequency variable dynamics is not practically easy, even if the delay coordinate of the observable variable is used.
To model high-frequency variable dynamics,
a method using a low-frequency variable is effective~\cite{nakai_2018}.
In this study, we first model the dynamics of a low-frequency variable using only the time series data of the variable and its delay coordinate.
Next, we construct a model that predicts the dynamics of the high-frequency variable from that of the low-frequency variable.
The set of the constructed models succeeds in describing the dynamics of the high-frequency variable.
This approach involving modeling via other variables
is effective for dealing with complex dynamics. 
Abarbanel et al.~\cite{abarbanel_1994} introduced a method to construct a function describing the targeted variables depending on the delay coordinate of another variable.
By applying the proposed method and adopting spatially localized functions, 
we construct ODEs that describe the targeted variable using mainly the delay coordinate of a different variable. 

We introduce a variant of the RfR method, 
which first models the simple coherent dynamics of a certain variable
using the RfR method, and this system is then used to model the complex and sometimes intermittent dynamics of a targeted variable.
The proposed method is referred to as the joint RfR method.
This method enables the construction of a system of ODEs describing the complex time series using only physically understandable variables,
and it dramatically enlarges the capabilities of methods for constructing data-driven ODEs.
We describe the joint RfR method and demonstrate its validity 
by applying it to two examples.
The first involves the time series of the R\"ossler equation, which is a well-known chaotic system.
The second concerns the time series of the energy function of a chaotic fluid flow.
The proposed method succeeds in
describing the connection between the two types of time series of a single dynamical system,
although their behaviors are quite different.

{\it Modeling Method.}
\begin{figure*}
    \centering
\includegraphics[width=2.0\columnwidth,height=1.0\columnwidth]{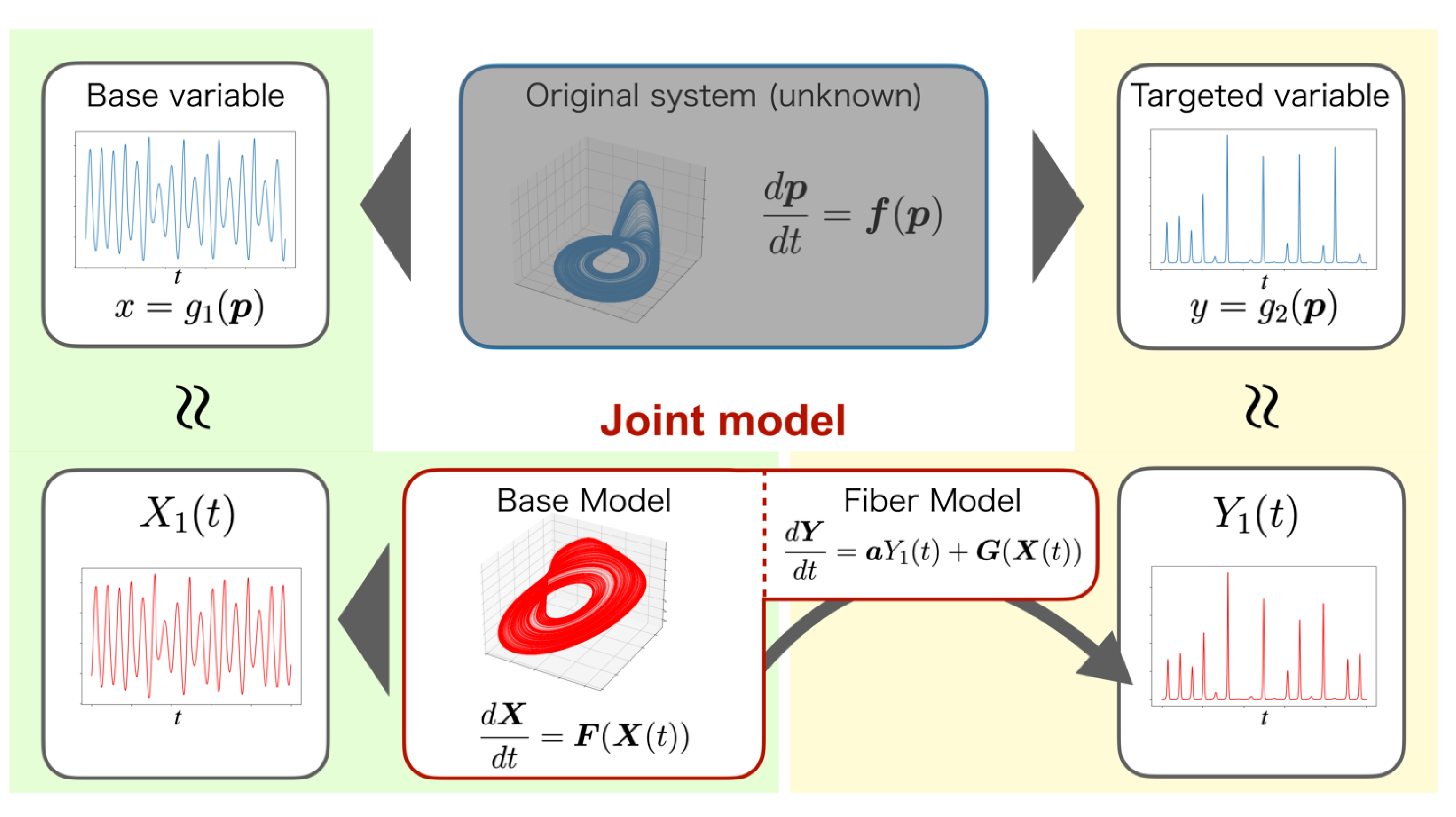}
    \caption{
        \textbf{Outline of the joint RfR method for constructing a joint model composed of the base and fiber models.} 
        The proposed method constructs a system of ODEs
        that infer the time series of two variables $x(t)$ and $y(t)$
        from only their observable time series data, 
        where $x(t)=g_1(\bm{p})$ and $y(t)=g_2(\bm{p})$ for some functions $g_1$ and $g_2$, and ${\bm{p}}$ is determined by some unknown dynamical system $\frac{d\bm{p}}{dt}=f(\bm{p}).$ 
       Using the time series of $x(t)$ 
        that shows relatively simple dynamics,
        we construct a ``base model'' 
        using a system of ODEs of a variable ${\bm X}$ composed of $x$ and its time delay variables
        employing
        the RfR method~\cite{tsutsumi22,tsutsumi23}. 
         The relatively complex behavior of the $y$ variable is described 
        as fiber dynamics using a system of skew-product-type ODEs of a variable ${\bm Y}$ composed of $y$, its time delay variables, and the
        variable ${\bm X}$ of the ``base model.'' 
        The constructed model for the fiber dynamics is called the ``fiber model''.
        Here the first component of the variables ${\bm X}$ and ${\bm Y}$ are $X_1$ and $Y_1$, respectively. 
 }
\label{fig:estimation-image}
\end{figure*}
We aim to achieve a data-driven construction of a system of ODEs for describing the relatively complex deterministic dynamics of a targeted variable. 
For this purpose, we assume that we can observe a time series of two variables showing simple and relatively complex dynamics.
\allblack
Because directly modeling the targeted variable's dynamics is difficult, we first construct a relatively simple dynamics model and use it to model the targeted variable's dynamics 
as the skew-product-type ODE, as described in figure~\ref{fig:estimation-image}. 
When the dimension of the delay coordinate for a variable showing relatively moderate dynamics is sufficiently high, 
invariant sets of the original system are reconstructed in the base model~\cite{takens_1981}.
The relatively complex dynamics of the original system are almost completely reconstructed by the base model.
The fiber model picks up the dynamics of the targeted variable from the base model.
The skew-product-type ODE is expected to demonstrate relatively complex dynamics.
We use the RfR method \cite{tsutsumi22} to estimate the base model describing 
the relatively simple variable~$x(t)$~(i.e., the base variable), 
and the new method is applied to modeling the relatively complex dynamics of a variable~$y(t)$~(i.e., the targeted variable) as a system of ODEs, which is referred to as the fiber model.
In this section, we briefly review the RfR method for constructing the base model 
and explain the method used for the targeted variable in detail.
Finally, a method to generate a numerical trajectory from the estimated system is explained. 
To describe the dynamics of a targeted observation $y(t)$, we construct a joint model composed of base and fiber models.\\
\indent{\bf Base model.} The first part of the joint model describes the dynamics of the base variable $x$. 
It is constructed from scalar time series data of $x(t)$ using the RfR method~\cite{tsutsumi22}.
In the RfR method,
the delay coordinate is employed to describe chaotic behavior using only a scalar observable time series. 
We estimate a $D$-dimensional system of ODEs of the variable 
${\bm X}=(X_1, X_2,\ldots, X_D)$, which almost satisfies the delay structures among components of the variable: 
\begin{align}
\label{eq:delay-structure_base}
    X_1(t) \approx X_2(t + \tau) \approx \cdots \approx X_D(t+(D-1)\tau),
 \end{align}
where $X_1$ is considered to describe the dynamics of $x$ and $\tau$ is the delay time.
See \cite{tsutsumi22} for details concerning the choice of $\tau$.
The model
\begin{align}
     \frac{d {\bm X}}{dt} = {\bm F}({\bm X}), \label{eq:BaseModel}
\end{align}
is constructed, where ${\bm F}({\bm X})$ is formulated by 
 \begin{equation}
    \label{eq:base-system_form}
    {\bm F}({\bm X}) = {\boldsymbol \beta}_0 + \sum_{d=1,\cdots,D} {\boldsymbol \beta}_d X_d + \sum_{j=1,\cdots,J} {\boldsymbol \beta}_{D+j}~\phi_j({\bm X}),
    \end{equation}
where 
    $\phi_j({\bm X}) = \exp \left( \frac{-||\bm{X}-c_j||^2}{\sigma^2} \right )$,
and the paramters $c_j (j=1,\cdots,J)$ 
are distributed as lattice points and $\sigma$ is the standard deviation of the Gaussian distribution. 
The model coefficients ${\boldsymbol \beta}_i (i=1,\cdots, D+J)$ are determined 
using the time derivative at each sample point estimated by the Taylor approximation. 
See \cite{tsutsumi22} for more details concerning this process.\\
\indent{It should be noted that 
if we can observe a time series of the base variable,
the base modeling process is not required, and a fiber model may be directly constructed to predict the time series of the targeted variable.
}\\
\indent{\bf Fiber model.} With the knowledge of the variable ${\bm X}$ obtained from the base model,
we next construct a model describing the dynamics of the targeted variable $y$.  
For a given delay time $\hat \tau(>0)$,
we construct a system of two-dimensional ODEs of the variable $\bm{Y} = (Y_1, Y_2)$, which is expected to satisfy the delay structures among components of the variable:
    $Y_1(t) \approx Y_2(t+\hat{\tau}).$ 
The relationship will be used for generating an appropriate numerical trajectory.
Using the variable ${\bm X}$ of the base model,
we form  
the ODEs describing ${\bm Y}$ 
as follows:
\begin{align}
\label{eq:fiber-model}
     \frac{d {\bm Y}}{dt} = {\boldsymbol a} Y_1 + {\bm G}({\bm X}), 
    \end{align}
where ${\bm G}({\bm X}) = {\boldsymbol \gamma}_0 + \displaystyle\sum_{d=1,\ldots,D} {\boldsymbol \gamma}_d X_d + \displaystyle\sum_{j=1,\ldots,J} {\boldsymbol \gamma}_{D+j}~\phi_j({\bm X}),$ 
and ${\boldsymbol \gamma}_i$ and ${\boldsymbol a}$ is the model coefficients 
and $\phi_j$ is the same as that in Eq.~\eqref{eq:base-system_form}, which means that $c_j$ of $\phi_j$ are distributed as lattice points for $\bm X$.
The model coefficients ${\boldsymbol \gamma}_i$ and ${\boldsymbol a}$ are determined via the time series data using ridge regression.
This formulation and the introduction of the linear term $Y_1$ into \eqref{eq:fiber-model} are important for modeling 
complex dynamics with restricted computational resources.
In the above formulation,
we can avoid distributing center points in the delay coordinate of the targeted variable.
See Supplemental Material 1 for the explanation of the linear term.
When we estimate the model coefficients,
we use the time series of the delay coordinate $(y(t), y(t-\hat{\tau}))$ as that of $\bm{Y}(t)$.
See \cite{tsutsumi22} for more details concerning this estimation. 
The appropriate delay time depends on the frequency of the time series, and the delay time $\tau$ for the base model and the delay time $\hat{\tau}$ for the fiber model need not be the same.
In our examples, $\hat{\tau}$ should be chosen to be smaller than $\tau$.\\
\indent {\bf Generation of a model trajectory.} Here, we introduce a method for estimating a numerical trajectory without detailed information for the variable $y$. 
In some data-driven models, 
an appropriate trajectory is not on its chaotic attractor but on its chaotic saddle.
A chaotic saddle is a chaotic invariant set that does not attract some neighborhoods. 
The bias and the shortage of learning data can easily cause the situation.
Even when a model trajectory is realized on a chaotic saddle, 
a short-term numerical trajectory via the forward integration of the model mimics that of the actual well,
but a long-term trajectory diverges.
In \cite{tsutsumi23}, instead of using the simple forward integration, we generate an appropriate numerical trajectory on a chaotic saddle of the 
model constructed using the RfR method
employing the Stagger-and-Step method~\cite{sweet_2001c}.
The appropriateness of a trajectory is determined by the delay structure~\eqref{eq:delay-structure_base} along a trajectory.
To generate a long trajectory 
of the base model, we employ the Stagger-and-Step method when the chaotic invariant set of the model is a saddle.
We also apply the idea of this Stagger-and-Step method to the given fiber model~\eqref{eq:fiber-model}.\\
\indent To generate a trajectory from the estimated fiber model~\eqref{eq:fiber-model},
The time series of the variable of the base model ${\bm X}(t)$ is required.
When the time series $\bm{X}(t)$ is given,
we can interpret the fiber model~\eqref{eq:fiber-model} as a linear system of ${\bm Y}$ with a time depending external force ${\bm G}({\bm X}(t))$.
Using the description of a solution of a linear system,
the solutions of the fiber model~\eqref{eq:fiber-model} can be analytically described.
\begin{table}[t]
\small
    \centering
    \begin{tabular}{|c|c|c|c|c|c|c|c|c|c|}
    \hline
        & $D$ & $\tau$  & $J$  & $\hat{\tau}$ & $t^*$ \\ \hline
        R\"ossler  &  3  & 0.60 & ~~~~7,716 & 0.40  & 0.60 \\ 
        Fluid flow &  7  & 2.00 &   832,291 & 1.00  & 1.50  \\ \hline
    \end{tabular}
    \caption{
    {\bf Sets of parameters.}
    These parameters are used for modeling the dynamics of the R\"ossler equation and fluid flow.
    \footnote{
    We set $(\delta_{grid},n,T,\Delta t, m, p) = (0.25,5\cdot 10^4,10^5,0.10,3,0.1)$ in the case of the R\"ossler equation, and $(\delta_{grid},n,T,\Delta t, m, p) = (0.40,5\cdot 10^4,5\cdot10^4,0.05,3,0.1)$ in the case of the fluid flow,  
    where the parameter $\delta_{grid}$ is the grid size used to determine the Gaussian radial basis function, 
    $n$ is the number of regression data points, 
    $T$ is the time length
    of the time series of the base and targeted variable, and $\Delta t$ is the time step of the time series data. 
    The parameter $\sigma$ of $\phi_j$
    is determined from $m$ and $p$
    (see \cite{tsutsumi23} for more details).
    }
    }
\label{table:parameters}
\end{table}
The appropriateness of a model trajectory is quantified by the delay structure $\Delta (t):= Y_1(t) - Y_2(t + \hat{\tau}),$
which is written as 
$
    \Delta (t) = C(t) Y_1(0) - Y_2(0) + M(t)
$,
where $C(t)$ and
$M(t)$ do not depend on $Y_1(0)$ and $Y_2(0)$.
Note that this form is a linear equation for $Y_1(0)$ and $Y_2(0)$.
The appropriate trajectory of the fiber model \eqref{eq:fiber-model} should satisfy $\Delta (t) \approx 0$ for any $t$.
Suppose $\Delta(0) = 0$ and $\Delta (t^*) = 0$~ for some given $t^*>0$, then $Y_1(0)$ and $Y_2(0)$ can be obtained only from the ODEs of ~ Eq.~\eqref{eq:fiber-model} (see Supplemental Material 2 for more detail).
By applying this approach, we can estimate a model trajectory 
corresponding to the targeted observation $y(t)$.\\
\indent The settings of the parameters for the example dynamics are given in Table~\ref{table:parameters}.
The following results can be obtained robustly with different sets of parameters.

{\it Example 1: R\"ossler equation.}
We model a system of differential equations using the time series of the R\"ossler equation~\cite{roessler_1976}: 
$
    \frac{dx}{dt} = -y -z, 
    \frac{dy}{dt} = x + 0.2y,
    \frac{dz}{dt} = 0.2 + xz -5.7z.
$
It is known that a $z$ variable behaves less coherently than $x$ and $y$ variables. Therefore, the direct modeling of $z$ dynamics is more difficult than that of $x$ or $y$ dynamics. 
As an example, we first model the dynamics of a variable $x$ by considering it to be the base variable and using it to model the $z$ dynamics.

\begin{figure}[t]
    \centering
\includegraphics[width=0.95\columnwidth,height=0.55\columnwidth]{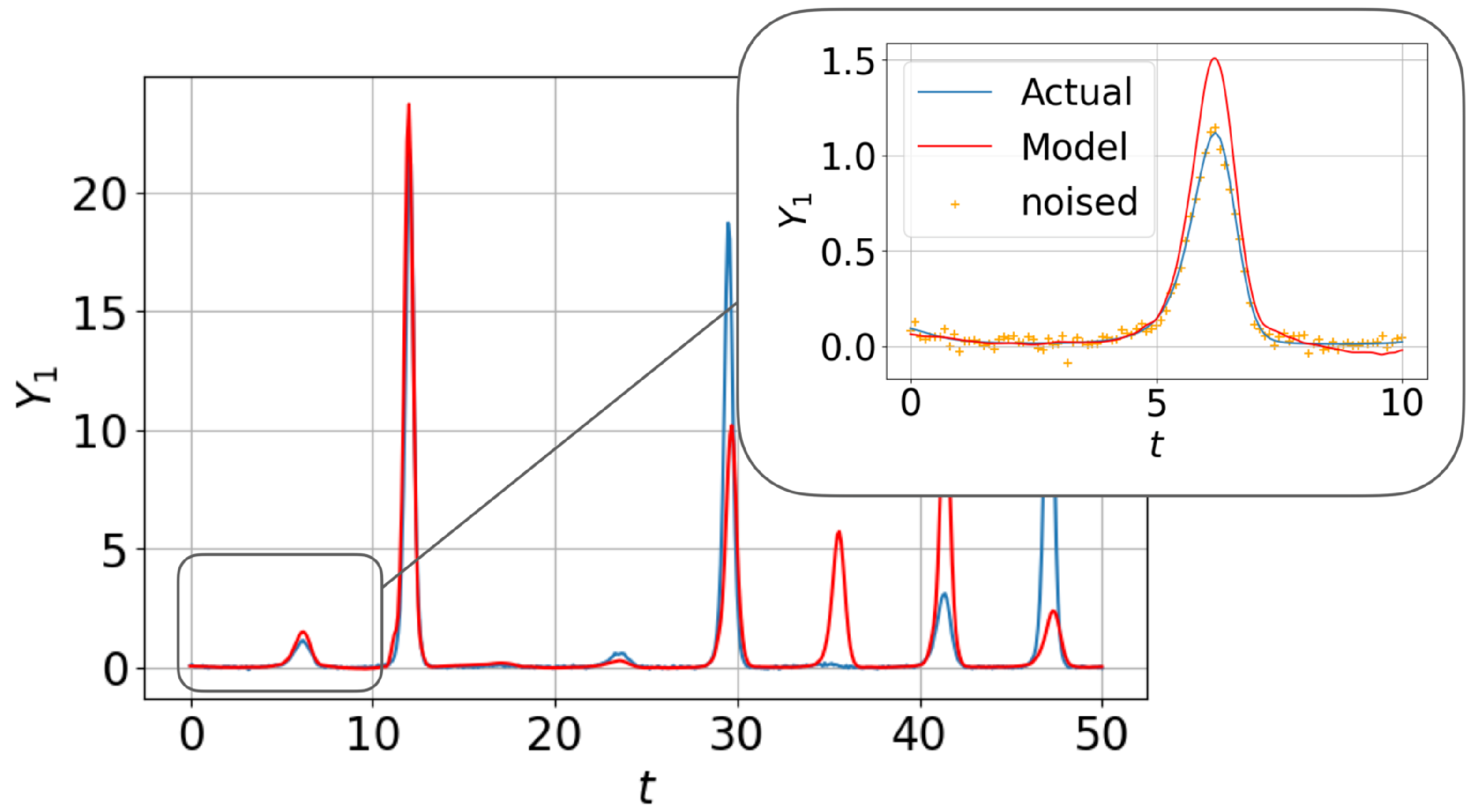}
    \caption{
        \textbf{Short time series for the R\"ossler dynamics.} 
        The estimated time series of {the targeted variable} $z$ ($Y_1$ of the fiber model)~(red) 
        is shown together with that of the actual time series~(blue).
        The time series including noise used for the estimation is also plotted~(yellow).
        We added Gaussian noise, with a standard deviation of 1\% of the raw data, to the time series of the $z$ variable.
    }
    \label{fig:rossler_short_summary}
\end{figure}

We assume that we can observe discrete time series of $x$ and $z$ variables with Gaussian noise whose standard deviation is $1\%$ of the standard deviation for each variable. 
See Fig.~\ref{fig:rossler_short_summary} for the time series of the $z$ variable with noise used for the modeling together with that without noise. 
{Here, the size of the noise is large enough to hide small amplitude fluctuations.}

\begin{figure}[t]
    \centering
     \includegraphics[width=0.95\columnwidth,height=0.55\columnwidth]{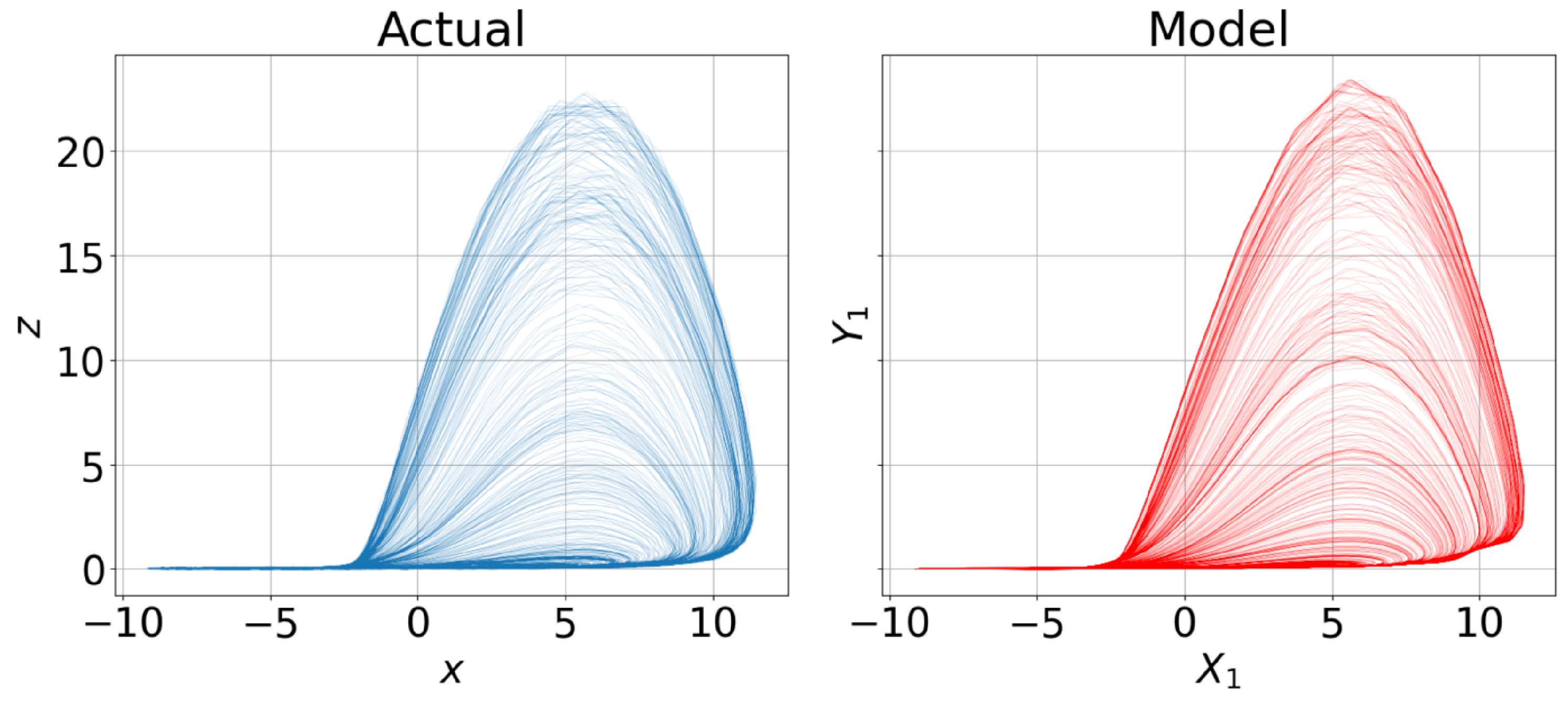}
    \caption{
        \textbf{Projection of a chaotic set for the R\"ossler dynamics.} 
        The projection of a long trajectory generated from 
 the original system onto the $(x,z)$ plane~(actual) is shown in the left panel and that of the model~(model) in the right panel.
        The two trajectories exhibit a similar shape.
    }
    \label{fig:rossler_whole-system}
\end{figure}

We construct a base model~\eqref{eq:BaseModel} for $\bm{X} :=(X_1, X_2, X_3)$ using the time series of the $x$ variable and a fiber model~\eqref{eq:fiber-model} for $\bm{Y} := (Y_1, Y_2)$ using the time series of $z$ together with the dynamics of the base model.
Refer to Fig.~\ref{fig:rossler_short_summary} for the short time series of the $Y_1$ variable obtained from the joint model, 
the skew-product-type model that joins the fiber and base models. 
It is shown that the joint model well predicts the short time series.
Figure~\ref{fig:rossler_whole-system} shows a projection of long trajectory points to a $(X_1, Y_1)$-plane.  
The results imply the successful reconstruction of the connection between the base and targeted variables.
Note that the positive Lyapunov exponent of the base model~($\lambda_1 = 0.07124$)
agree with that of the original R\"ossler equation~($\lambda_1 = 0.07123$), 
and the second Lyapunov exponents~$\lambda_2$ of both systems are neutral.
The negative Lyapunov exponent of the model is $-0.27506$,
and 
that of the original is $-5.21974$.
This discrepancy of the negative Lyapunov exponents is commonly observed in data-driven modelings~\cite{tsutsumi22,hart2024}.
%

\begin{figure}[t]
    \centering
\includegraphics[width=0.95\columnwidth,height=0.55\columnwidth]{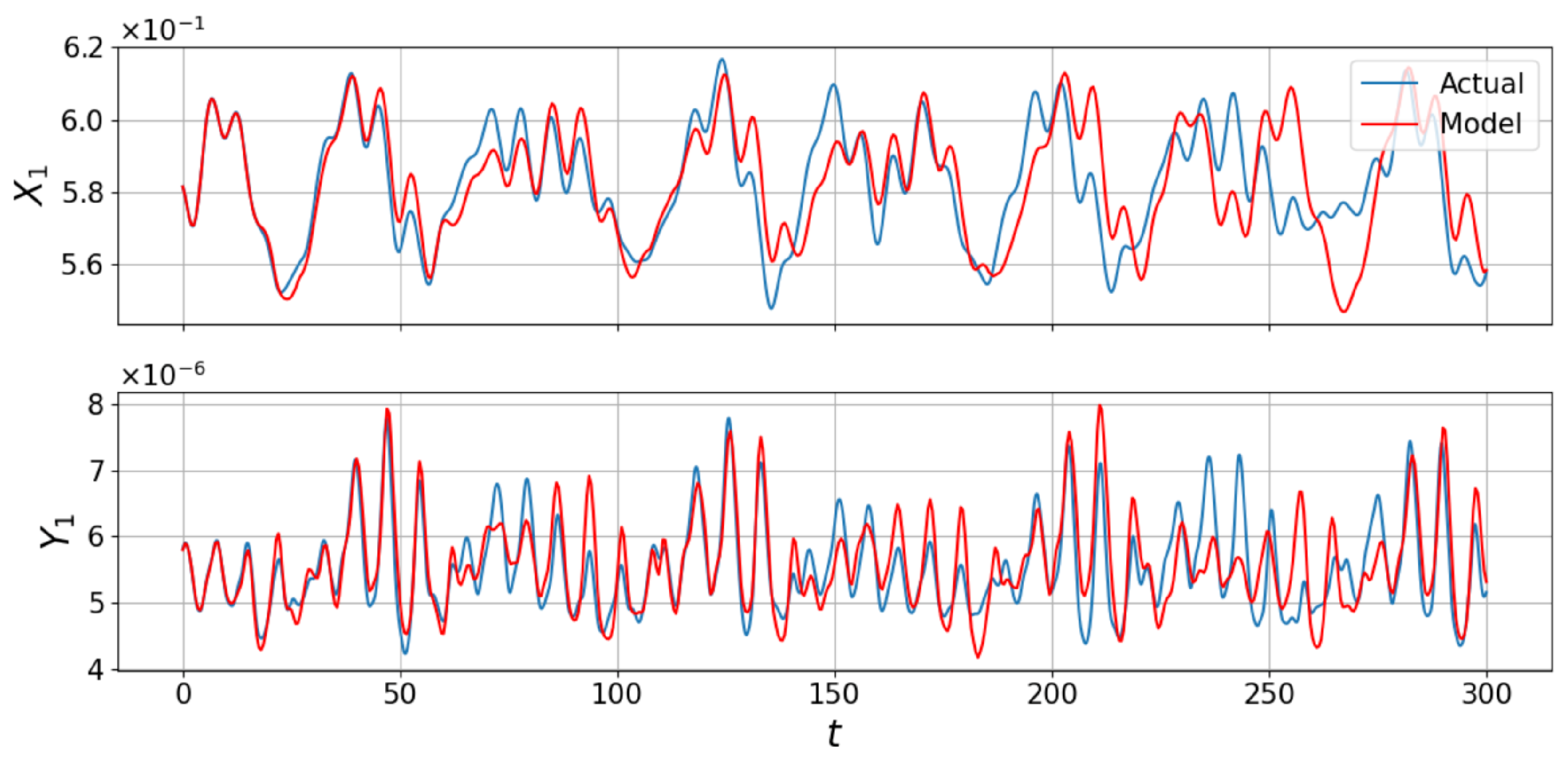}
    \caption{
        \textbf{Short time series of two different variables for the macroscopic behavior of the fluid dynamics.
        } 
        {
        The relatively moderate estimated time series of the base variable $E_1$ ($X_1$ of the base model)~(red) is shown in the upper panel. The relatively complex estimated time series of the targeted variable $E_{10}$ ($Y_1$ of the fiber model)~(red) is shown in the lower panel.
        }
        The actual trajectories are plotted together with the estimations, which are shown by the blue lines.
    }    \label{fig:fluid_short_sas-base}
\end{figure}

\begin{figure}[t]
    \centering    \includegraphics[width=0.95\columnwidth,height=0.6\columnwidth]{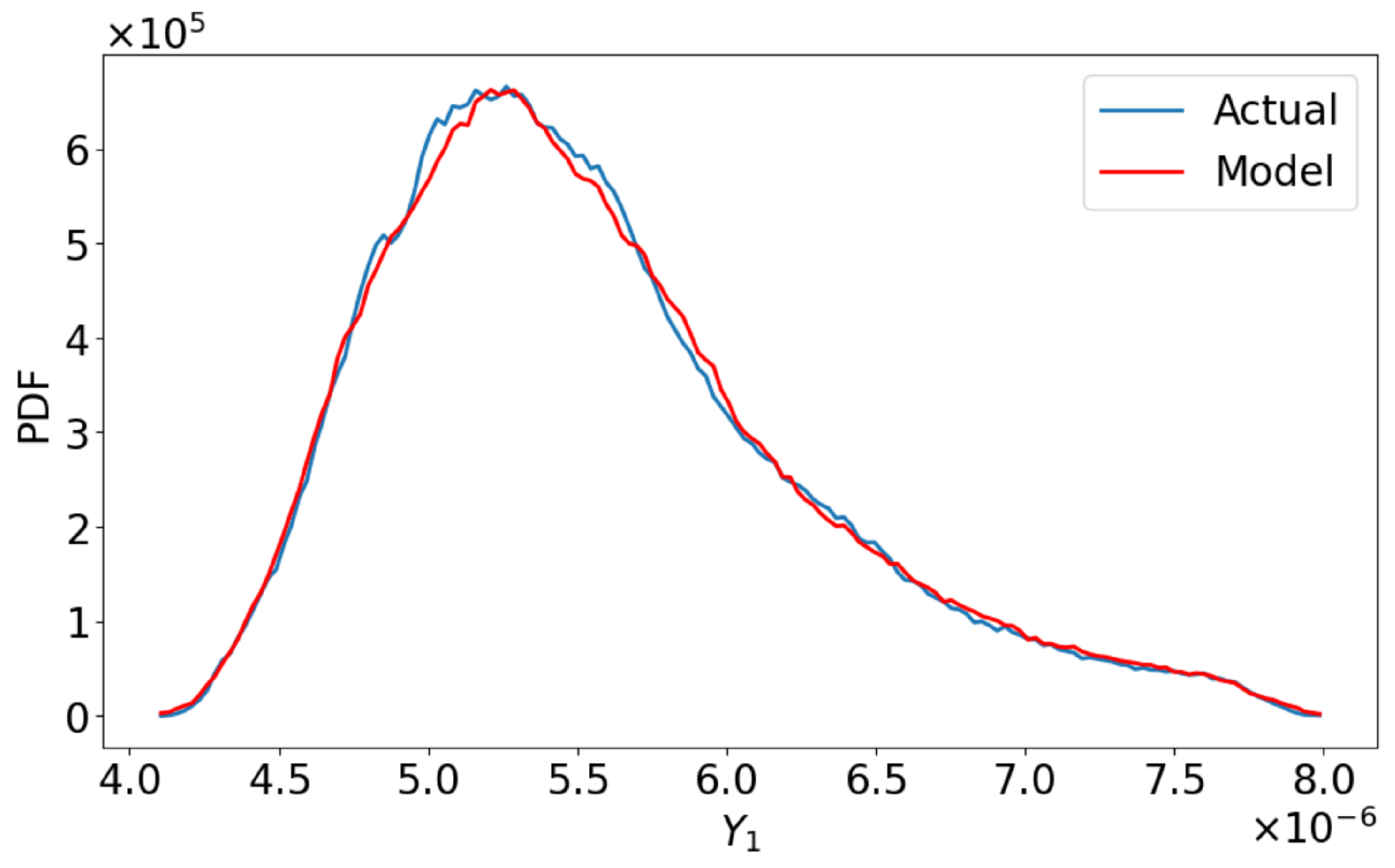}
    \caption{
        \textbf{ Density distribution of a targeted variable $Y_1$ for the macroscopic behavior of the fluid dynamics.} 
        The density distribution created from a long model trajectory ($T=50,000$) of $Y_1$ (estimation of $E_{10}$)~(red) is shown together with the actual trajectory~(blue). 
        The distributions exhibit similar shapes.
    }
    \label{fig:fluid_density}
\end{figure}

{\it Example 2: Fluid system.}
Next, we model the complex dynamics of a high-frequency fluid variable using a model of the simpler dynamics of a low-frequency variable.
It is known that the high-frequency dynamics of fluid flow show intermittency and complex behavior.
To obtain time series data for training, we consider the direct numerical simulation of the incompressible three-dimensional Navier--Stokes equation under periodic boundary conditions,  
and the forcing is input into the low-frequency variables at each time step so as to preserve the energy of the low-frequency part. 
{The settings of the numerical simulation are the same as those used by \cite{nakai_2018} with a viscosity parameter of $0.0585$.}
We exemplify a time series of energy functions $E_k(t)$ for wavenumber $k \in \mathbb{N}$, 
which is defined by 
$
    E_k(t) := \frac{1}{2}\sum_{D_k} \sum_{\zeta=1}^{3} |\mathcal{F}_{[v_{\zeta}]}(\kappa,t)|^2,
$
where $D_k:=\{ \kappa \in \mathbb{Z}^3 | k -0.5 \leq |\kappa| <k +0.5 \}$, $v_i$ is the $i$th component of the velocity, and $\mathcal{F}_{[v_i]}$ is the Fourier coefficient of the variable $v_i$.
We assume that the time series of the
$E_1(t)$ and $E_{10}(t)$ variables is known. 
In this section, we construct a base model (\ref{eq:BaseModel}) for $\bm{X} := (X_1, \ldots, X_7)$ using the time series of the $E_1(t)$ variable and a fiber model \eqref{eq:fiber-model} for $\bm{Y} := (Y_1, Y_2)$ using the time series of the $E_{10}(t)$ variable. 
Note that the turbulent dynamics of the high-wavenumber variable $E_{10}(t)$ 
is much more complex than that of the low-wavenumber variable $E_1(t)$.
We succeeded in achieving the short-term prediction of the $E_1(t)$ variable using the base model (Fig.~\ref{fig:fluid_short_sas-base} (top)) and $E_{10}(t)$ variable using the joint model (Fig.~\ref{fig:fluid_short_sas-base} (bottom)).
Although the long time series prediction failed 
because of the chaotic property, 
the density distribution of $E_{10}$ can be reproduced from the time series of the model (Fig.~\ref{fig:fluid_density}).
These results imply that the constructed joint model can well describe the dynamics of $E_{10}$.

\begin{figure}
    \centering \includegraphics[width=0.95\columnwidth,height=0.5\columnwidth]{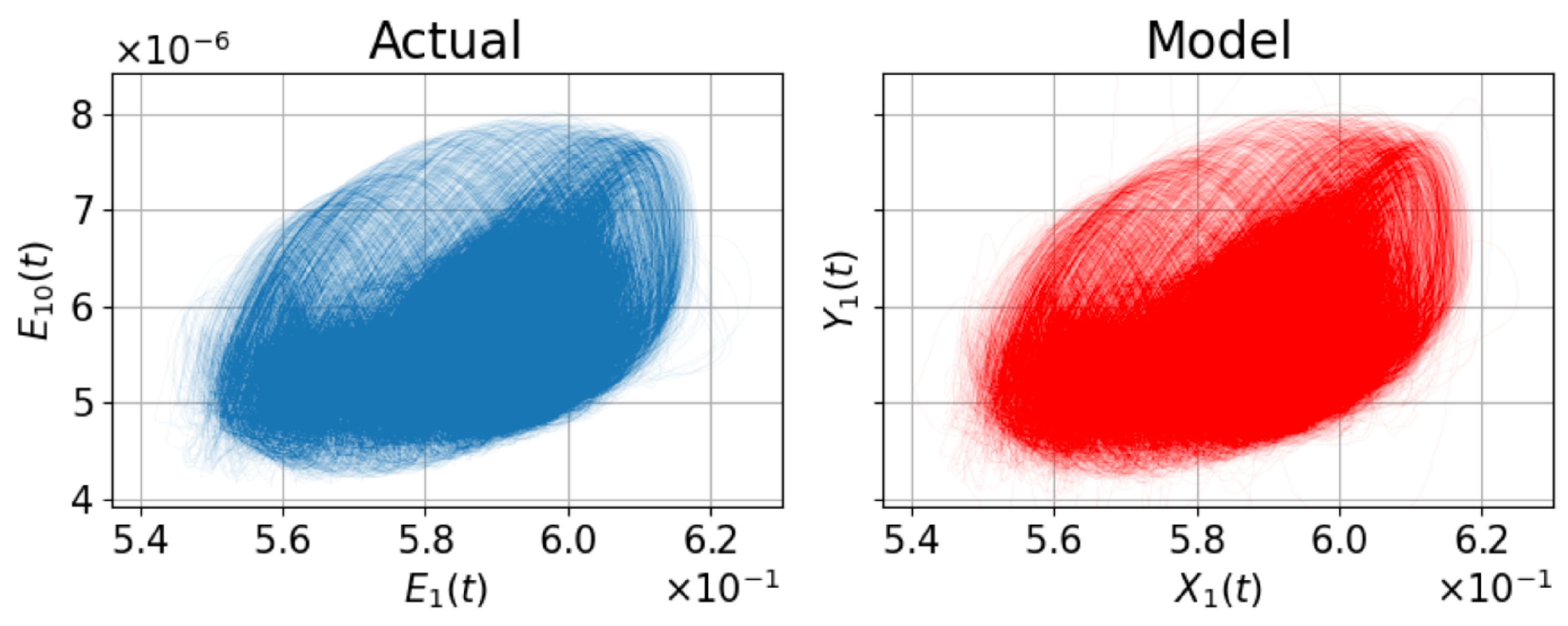}
    \caption{
        \textbf{Projection of a chaotic set for the macroscopic behavior of the fluid dynamics.} 
        The projection of a long trajectory generated from the original system onto the $(E_1, E_{10})$ coordinate~(actual) is shown in the left panel and that of the model~(model) is shown in the right panel.
        These two systems exhibit similar shapes.
        Note that the original system has an attractor, whereas the corresponding chaotic set in the model is a saddle.
    }
    \label{fig:fluid_joint-plot}
\end{figure}

The joint model reconstructs the connection between $E_1$ and $E_{10}$~(Fig.~\ref{fig:fluid_joint-plot}).
This imitation of the connection between the two energy variables contributes to the success of describing the high-frequency variable, which is difficult to model.
The positive Lyapunov exponents of the base model are $\lambda_1 = 0.2157, \lambda_2=0.0718, \lambda_3=0.0200$,
although we do not have the actual exponents for comparison.
We cannot obtain the Lyapunov exponents using a governing equation,
because a closed-form equation for the energy variable cannot be derived from the Navier--Stokes equations.
Since the attractor dimension is high,
it is hard to estimate Lyapunov exponents directly from time series~\cite{wolf1985}.

\newpage

{\it Concluding remarks.} 
The proposed method of constructing a system of ODEs from an observable time series is useful for modeling complex and multiscale dynamics in time and space physically only using understandable variables without background knowledge.
The computational cost to model behaviors of two or more variables with different time scales is high.
In such a situation, 
the effective strategy that we employ is to model moderate dynamics and use it to model complex dynamics through a fiber model. 
The computational cost to model moderate dynamics with the delay coordinate is lower,
and describing the major part of the complex dynamics with the variable of the base model reduces the required computational resource.
This skew-product-type modeling succeeds 
because the invariant sets of the original system are reconstructed in the base model
when the dimension of the delay coordinate is sufficiently high.
The developed method will be used to model multiscale dynamics, such as those of weather, beginning with the construction of a base model of low-frequency dynamics.

{\it Acknowledgement.} 
YS was supported by JSPS KAKENHI Grant Nos. 19KK0067, 21K18584, 23H04465, and 24K00537.  
KN was supported by JSPS KAKENHI Grant No.22K17965. 
The computations were carried out using the JHPCN (jh230028 and jh240051) and the Collaborative Research Program for Young $\cdot$ Women Scientists of ACCMS and IIMC, Kyoto University. 
%


\clearpage
\section*{Supplemental Material 1: Necessity of the linear term}
\label{sec:sup-material1}
In this section,
we explain the reasons why the linear term of $Y_1$ in the fiber model~\eqref{eq:fiber-model} is necessary to construct a model practically.
When the dimension of the base model $D$ is high, 
the dynamics of the targeted variable
may be described without a linear term as follows:
\begin{align}
    \label{eq:fiber-model-without-linear}
    \frac{d \bm{y}}{dt} = \hat{\bm{G}}(\bm{X}),
\end{align}
where $\bm{y}(t) := (y(t), y(t-\hat{\tau}))$ and $y(t)$ is the targeted variable.
To construct this system using numerical data,
the constructed model includes the estimation and numerical errors, which means that the estimated model can be described as 
\begin{align}
    \label{eq:fiber-model-with-noise}
    \frac{d \bm{Y}}{dt} = \hat{\bm{G}}(\bm{X}) + \bm{\delta}(\bm{X}).
\end{align}
By using an adequate modeling method and sufficient data, $\bm{\delta}(\bm{X})$ is determined to be small.  
A trajectory of the model~\eqref{eq:fiber-model-with-noise}
can be described as
\begin{align*}
    \bm{Y}(t) = \bm{y}(t) +\int_{0}^{t} \bm{\delta}(\bm{X}(s)) ds.
\end{align*}
This implies that a long-generated trajectory includes accumulated errors, and the effect of these errors gradually increases.
This deviation is caused by the independence of the right-hand side of the fiber model~\eqref{eq:fiber-model-with-noise} from $\bm{Y}$.
In the joint RfR method, the linear term of $Y_1$ is employed to avoid this issue.

\section*{Supplemental Material 2: Generation of a model trajectory from the fiber model}
\label{sec:sup-material2}
Here, we explain the method used to generate a model trajectory of the fiber model in detail.
To simplify the descriptions, 
we denote $\bm{G} (\bm{X}(t))$ of the fiber model~\eqref{eq:fiber-model} as $\bm{G}(t)$ and the $i$-th component of the vector $\bm{G}(t)$ and $\bm{a}$ of the fiber model~\eqref{eq:fiber-model} as $G_i(t)$ and $a_i$, respectively.
The estimated fiber model~\eqref{eq:fiber-model} can be described as follows:
\begin{align*}
    \frac{d {\bm Y}}{dt} = {\boldsymbol a} Y_1 + {\bm G}(t),
\end{align*}
which is a linear system of ${\bm Y}$.
The solution of this linear system can be described as
\begin{align*}
    & Y_1(t) = Y_1(0)~e^{a_1 t} + e^{a_1 t}~K(t),\\
    & Y_2(t) = Y_2(0) + \frac{a_2}{a_1} \left(e^{a_1 t} - 1 \right)Y_1(0) + \frac{a_2}{a_1}e^{a_1 t} K(t)+ \frac{1}{a_1} L(t),
\end{align*}
where 
\begin{align*}
    K(t) &= \int_0^t e^{-a_1 s}~G_1(s)ds,\\ 
    L(t) &= \int_0^t a_1 G_2(s) - a_2 G_1(s)ds.
\end{align*}
The appropriateness of a model trajectory 
is quantified by the delay structure $\Delta (t):= Y_1(t) - Y_2(t + \hat{\tau}),$
which is written as 
follows using the above forms:
\begin{align*}
    \Delta (t) = C(t) Y_1(0) - Y_2(0) + M(t),
\end{align*}
where
\begin{align*}
    C(t) &= \left(1 - \frac{a_2}{a_1} e^{a_1 \hat{\tau}}\right)e^{a_1 t} + \frac{a_2}{a_1} , \\
    M(t) &= e^{a_1 t}\left(K(t) - \frac{a_2}{a_1} e^{a_1 \hat{\tau}} K(t+\hat{\tau})\right) - \frac{1}{a_1} L(t+\hat{\tau}).
\end{align*}
Note that this form is a linear equation for $Y_1(0)$ and $Y_2(0)$.
The appropriate trajectory of the fiber model \eqref{eq:fiber-model} should satisfy $\Delta (t) \approx 0$ for any $t$.
Suppose $\Delta(0) = 0$ and $\Delta (t^*) = 0$~ for some given $t^*>0$; 
$Y_1(0)$ and $Y_2(0)$ can then be obtained using only the fiber model~\eqref{eq:fiber-model} as follows:
\begin{align*}
    Y_1(0) &= - \frac{M(t^*) - M(0)}{C(t^*) - C(0)},\\
    Y_2(0) &= \frac{C(t^*) M(0) - C(0) M(t^*) }{C(t^*) - C(0)}.
\end{align*}
By applying this approach, 
we can estimate a 
model trajectory 
corresponding to the targeted observation $y(t)$.
\end{document}